\documentclass[9pt,twocolumn,twoside]{opticajnl}
\journal{opticajournal} 

\setboolean{shortarticle}{true}

\title{Revealing the propagation dynamic of Laguerre-Gaussian beam with two Bohm-like theories}

\author[1,2]{Peng-Fei Huang}
\author[1,2,*]{Ya Xiao}
\author[1]{Shan-Chuan Dong}
\author[1]{Yong-Jian Gu}

\affil[1]{College of Physics and Optoelectronic Engineering, Ocean University of China, Qingdao, 266100, People's Republic of China}
\affil[2]{These authors contributed equally to this work}

\affil[*]{xiaoya@ouc.edu.cn}

\begin{abstract}
By employing $x$-Bohm theory and $p$-Bohm theory, we construct the position and momentum trajectories of single-mode and superposed-mode Laguerre-Gaussian (LG) beams. The dependence of divergence velocity and rotation velocity on the initial position and propagation distance is quantified, indicating that LG beams exhibit subluminal effects, even in free space. Additionally, we clarify the formation of the petal-shaped intensity distribution of the superposed-mode LG beam in terms of motion trajectory, where the particle-like trajectory and wave-like interference  are ``simultaneously" observed. Our work provides an intuitive way to visualize the propagation characteristics of LG beams and deepen the comprehension of Bohm-like theory.
\end{abstract}

\setboolean{displaycopyright}{false} 

\begin{document}

\maketitle

\section{Introduction}
Laguerre-Gaussian (LG) beams, a typical type of vortex beams carrying orbital angular momentum (OAM), are renowned for the infinite OAM orthogonal modes and spiral phase structure \cite{ABS1992}. They have been widely used in optical communication \cite{WPS2021,WHY2015,BYR2013}, particle manipulation \cite{PB2011, GGR2018,OD2020}, super-resolution imaging \cite{SZ2023,WGZ2023, KMS2018}, quantum information processing \cite{W2015,FRG2020, PNO2020,KZZ2023}, and various other domains \cite{PFS2013,L2013,HSF2022}. These promising applications of LG beams have driven significant advancements in their generation and detection techniques \cite{ZWY2020,LQW2022,WL2022,XLS2022,ZZW2021}.  Moreover, several adaptive optical methods have been devised to recover LG beams distorted by atmospheric or ocean turbulence \cite{DW2017,YCC2020}. Despite all these works, investigations into the mechanisms and physics underlying their propagation remain limited. Most existing studies have approached this subject from a classical wave-optics analysis \cite{LRK2017,CWL2022}. Further research is needed from a trajectory viewpoint, considering the wave-particle duality, to gain insight into the propagation characteristics inherent in different LG beams.  For instance, it is crucial to investigate why LG beams  exhibit slow-light effects, even in free space \cite{BHM2016,LRW2018}; determine the quantitative relationship between beam parameters and propagation velocity; and elucidate how the petal-shaped intensity distribution of superposed-mode LG beams forms and evolves.

In standard quantum mechanics, the Heisenberg uncertainty principle suggests that it is impossible to simultaneously determine the position and momentum of a particle \cite{H1927}. This implies that accurately tracking the trajectory of a quantum particle, as we can with a classical particle, is unattainable. Nevertheless, in 1952, Bohm proposed a nonlocal hidden-variable interpretation of quantum theory, predicting that a particle has a deterministic trajectory guided by the position space wave function \cite{B1952,Bohm1952}. In Bohm's interpretation, the position variable is regarded as the primary ontological variable. The momentum of a particle is simply derived from its velocity, which is determined by the gradient of the wavefunction at the corresponding position. Considering the equal importance of position and momentum inherent in both classical and quantum mechanics, Epstein later proposed an alternative trajectory theory, where momentum acts as the primary ontological status and a particle's trajectory is fundamentally characterized by its momentum space wavefunction \cite{E1953,Epstein1953}.  For simplicity, we refer to Bohm's original theory as $x$-Bohm theory and Epstein's proposal as $p$-Bohm theory. Subsequently, Wiseman demonstrated that velocities of Bohm-like particles can be experimentally extracted by taking conditional averages of weak measurements on an ensemble of post-selected systems \cite {W2007}. Inspired by his idea, researchers have successfully obtained putative Bohmian trajectories in various two-slit experiments \cite{KBR2011,MRF2016,XKX2017,XWX2019,PFG2020}. However, the propagation dynamics of LG beams, particularly that of superposed-mode LG beams, have been rarely analyzed from the perspective of Bohm-like trajectories. 

In this work, we explore the propagation trajectories of single-mode and superposed-mode LG beams within the frameworks of both $x$-Bohm theory and $p$-Bohm theory. We investigate the subluminal effect of these beams concerning the initial position and the propagation distance. Additionally, the formation of petal-shaped intensity distributions of superposed-mode LG beams is clarified by the motion trajectory.  Notably, we find that both the position trajectory and the momentum trajectory exhibit significantly different dynamics in $x$-Bohm theory and $p$-Bohm theory. 
Our trajectory-baesd method not only provides an intuitive depiction of the propagation characteristics of LG beams but can also be extended to describe the properties of arbitrarily structured light beams, including propagation behavior, spin-orbit and orbit-orbit interactions, as well as diffraction and interference phenomena. 
Additionally, by employing the weak measurement techniques mentioned in the references \cite{PFG2020, YXL2020}, it can also experimentally reconstruct the corresponding trajectories.

\section{Theoretical Model}
We start by reviewing the $x$-Bohm theory and $p$-Bohm theory, and then derive the trajectories of the primary and non-primary ontological variables of the particles in each theory. This enables us to quantify the difference between the two theories by contrasting their predicted trajectories in the same space.

\subsection{the trajectory in $x$-Bohm theory}\label{subsec2}

In $x$-Bohm theory, the primary and non-primary ontological variables are respectively identified as the position and momentum of the particle,  with the fundamental dynamics occurring in position space \cite{B1952,Bohm1952}. Assuming the particle's position wave function is $\psi(\textbf{x},t)=R(\textbf{x},t){\rm exp}(iS(\textbf{x},t)/\hbar)$, its velocity is given by

\begin{equation}\label{xve0}
	\textbf{v}(\textbf{x},t)=\nabla S(\textbf{x},t)/m,
\end{equation}
where $\nabla $ is the gradient operator in position space, $m$ is the mass of the particle, and $ R(\textbf{x}(t),t) $ and $ S(\textbf{x}(t),t) $ are the amplitude and phase functions, respectively. As the wave function $\psi(\textbf{x},t)$ evolves according to the Schr\"odinger equation
\begin{equation}\label{xs-equ}
	i\hbar \dfrac{\partial}{\partial t} \psi(\textbf{x},t)= \dfrac{\hat{\textbf{p}}^2}{2m}\psi(\textbf{x},t)+\hat{V}(\textbf{x}) \psi(\textbf{x},t).
\end{equation}
Here, $\hat{\textbf{p}}=-i\hbar \nabla$ is the momentum operator. Eq. (\ref{xve}) can be rewritten as
\begin{equation}\label{xve}
	\textbf{v}(\textbf{x},t)=\dfrac{\textbf{j}(\textbf{x},t))}{\vert \psi(\textbf{x},t))\vert^{2}},
\end{equation}
where $ \textbf{j}(\textbf{x},t)$ represents the probability current and can be expressed as 
\begin{equation}\label{xcurrent}
	\begin{split}
		\textbf{j}(\textbf{x},t) = \dfrac{1}{2m}(\psi^{\ast}(\textbf{x},t) \hat{\textbf{p}} \psi(\textbf{x},t)-\psi(\textbf{x},t) \hat{\textbf{p}}  \psi^{\ast}(\textbf{x},t)).
	\end{split}
\end{equation}
In analogy to classical Newtonian mechanics, the position trajectory of a particle in the $x$-Bohm  theory is determined by its initial position $ \textbf{x}_{0} $ and the integral of its velocity over time. The guidance equation for the position trajectory can be written as
\begin{equation}\label{xxtra}
	\begin{split}
		\textbf{x}(t) = \textbf{x}_{0} + \int^{t}_{0} \textbf{v}(\textbf{x},t)dt.
	\end{split}
\end{equation}
Clearly, for a particle at position $ \textbf{x}(t) $ at time $ t$, the momentum of the particle can be obtained by multiplying its velocity by its mass, i.e., 
\begin{equation}\label{xptra}
	\begin{split}
		\textbf{p}(\textbf{x}(t),t) = m\textbf{v}(\textbf{x}(t),t).
	\end{split}
\end{equation}
From an experimental perspective, Wiseman et al. have shown that  Eq. (\ref{xptra}) can be expressed in terms of the weak value as follows:
\begin{equation}\label{xptraj}
	\begin{split}
		\textbf{p}(\textbf{x}(t),t)={\rm Re}\dfrac{\langle \textbf{x}(t)\vert \hat{\textbf{p}}\vert \psi(t)\rangle}{\langle \textbf{x}(t)\vert  \psi(t)\rangle}= \hbar\cdot {\rm Im }\dfrac{ \nabla \psi(\textbf{x}(t),t) }{\psi(\textbf{x}(t),t)}.\\
	\end{split}
\end{equation}
By tracking $ \textbf{p}(\textbf{x}(t),t) $ as $ t $ evolves, one can assign a momentum trajectory to the particle in the $x$-Bohm theory.

\subsection{the trajectory in $p$-Bohm theory}\label{subsec2}

The $p$-Bohm theory offers a different perspective on the dynamic behavior of a particle, wherein the momentum and position of the particle are identified as the primary and non-primary ontological variables, respectively \cite{E1953,Epstein1953}. Clearly, its fundamental dynamics take place in momentum space. The particle's wave function in the momentum representation can be expressed as $\widetilde{\psi}(\textbf{p},t) =\int^{\infty}_{-\infty}d^{3}x\psi(\textbf{x},t)e^{-i\textbf{p}\cdot\textbf{x}/\hbar}$  and is guided by the Schr\"odinger equation 
\begin{equation}\label{ps-equ}
	i\hbar \dfrac{\partial}{\partial t} \widetilde{\psi}(\textbf{p},t)= \dfrac{\hat{\textbf{p}}^2}{2m}\widetilde{\psi}(\textbf{p},t)+\hat{V}(i\hbar\nabla_{p} \widetilde{\psi}(\textbf{p},t)),
\end{equation}
where $\nabla_{p}$ is the gradient operator in momentum space.  Unlike  Eq. (\ref{xs-equ}) involves only second derivatives with respect to $x$, Eq. (\ref{ps-equ}) can involve any number of derivatives with respect to $p$, potentially giving rise to distinct trajectories in $p$-Bohm theory.

Similarly, the evolution of the particle's momentum trajectory is determined by the guidance equation  
\begin{equation}\label{pptra}
	\begin{split}
		\textbf{p} (t) = \textbf {p}_{0} + \int^{t}_{0} \textbf{v}(\textbf{p},t)dt=  \textbf {p}_{0} + \int^{t}_{0} \dfrac{\textbf{j}(\textbf{p},t)}{\vert \widetilde{\psi}(\textbf{p},t)\vert^{2}}dt,
	\end{split}
\end{equation}
where $ \textbf {p}_{0} $ is the initial momentum, $ \textbf{v}(\textbf{p},t)=\textbf{j}(\textbf{p},t)/\vert \widetilde{\psi}(\textbf{p},t)\vert^{2} $ is the corresponding velocity, and $ \textbf{j}(\textbf{p},t) $  is the probability current, writing 
\begin{equation}\label{pcurrent}
	\begin{split}
		\textbf{j}(\textbf{p},t)=\dfrac{2}{\hbar}\int^{p}_{-\infty} {\rm Im}[\widetilde{\psi}(\textbf{p}',t)^{\ast}V(i\hbar\nabla_{p'})\widetilde{\psi}(\textbf{p}',t)]d^{3}p' .
	\end{split}
\end{equation}
It satisfies the physically reasonable assumption that $\textbf{j}(\textbf{p},t) \rightarrow 0$ as $ \vert \textbf{p} \vert \rightarrow \infty$.

Analogous to Eq. (\ref{xptraj}), for a particle with momentum $\textbf{p}(t) $ at time $ t$, the position of the particle can be expressed as
\begin{equation}\label{pxtra}
	\begin{split}
		\textbf{x}(\textbf{p}(t),t)={\rm Re}\dfrac{\langle \textbf{p}(t)\vert \hat{\textbf{x}}\vert \widetilde{\psi}(t)\rangle}{\langle \textbf{p}(t)\vert  \widetilde{\psi}(t)\rangle}= -\hbar\cdot {\rm Im }\dfrac{ \nabla_{p} \widetilde{\psi}(\textbf{p}(t),t) }{\widetilde{\psi}(\textbf{p}(t),t)}.\\
	\end{split}
\end{equation}
By tracking $ \textbf{x}(\textbf{p}(t),t) $ as $t$  evolves, one can assign a position trajectory to the particle in the $ p$-Bohm theory.

It is noteworthy that the energy flow trajectories \cite{LRK2017, CWL2022}, employed in previous investigations on the propagation of LG beams, are essentially similar to the position trajectories in position space. By combining $x$-Bohm theory with $p$-Bohm theory, an in-depth exploration of the position trajectory in momentum space, along with the momentum trajectory in both position and momentum spaces, becomes feasible. Subsequently, a comprehensive analysis of LG beam propagation trajectory in three-dimensional space will be conducted to achieve a more thorough understanding of its dynamics.

\section{the trajectory of LG Beam}\label{sec3}

In position space, the time-dependent wave function (or complex amplitude at the propagation distance $z$) of a single-mode LG beam with an angular momentum index $l$ and a radial mode index $p$ can be expressed as \cite{ABS1992}

\begin{equation}\label{xLG}
	\begin{split}
		LG^{ l}_{p}(r,{\theta},z)& =\sqrt{\dfrac{2p!}{\pi(p+\vert l\vert)!}}\frac{1}{\omega_{z}}(\dfrac{\sqrt{2}r}{\omega_{z}})^{\vert l \vert}{\rm exp}(\dfrac{-r^{2}}{\omega_{z}^{2}})\times\\
		& L^{\vert l \vert}_{p}(\dfrac{2r^{2}}{\omega_{z}^{2}}){\rm exp}(-\dfrac{ikr^{2}z}{2(z^{2}+z^{2}_{R})}-il{\theta} +i\xi_g +ikz),
	\end{split}
\end{equation}
where $(r, \theta)$ are the polar coordinates, $k =2\pi/\lambda$ is the wave number, $\lambda $ is the wavelength, $z_{R}=\pi\omega_{0}^{2}/\lambda$ is the Rayleigh length, $\omega_{z}=\omega_{0}\sqrt{1+(z/z_{R})^{2}}$ is the beam waist radius at the propagation distance $z$, $L^{\vert l \vert}_{p}(\cdot)$ the associated Laguerre polynomial, $\xi_g= (2p + \vert l \vert + 1){\rm arctan}({z}/z_{R} )$ is the Gouy phase, and ${\rm exp}(il\theta)$ is the spiral phase term. For simplicity, we consider an LG beam with a radial mode index $p=0$. The wave function of the LG beam in momentum space is obtained by applying the Fourier transform to its position space wave function, which can be expressed as   
\begin{equation}\label{pLG}
	\begin{split}
		\widetilde{LG}^{l}_{0}(\rho,\varphi,z)& =\sqrt{\dfrac{2\pi}{\vert l\vert!}} \omega_{0}(-i)^{\vert l\vert} (\sqrt{2}\pi\omega_{0}\rho)^{\vert l \vert}\times\\
		&{\rm exp}[-\pi^{2}\rho^{2}\omega_{0}^{2}(1+i  \dfrac{z}{z_{R}})] {\rm exp}(-il\varphi + {ikz}),
	\end{split}
\end{equation}
where $(\rho,\varphi)$ are the polar coordinates. The position space wave function $ \psi(r,\theta,z) $ and momentum space wave function $ \widetilde{ \psi}(\rho,\varphi,z) $ of a superposition state of two LG beams with different angular momentum indices $l_{1}$ and $l_{2}$, and same radial indices $p_{1}=p_{2}=0$, can be respectively expressed as  
\begin{equation}\label{spLG}
	\begin{split}
		& \psi(r,\theta,z) = a_{1} LG^{l_{1}}_{0}(r,\theta,z)+a_{2} LG^{ l_{2}}_{0}(r,\theta,z),\\
		&\widetilde{ \psi}(\rho,\varphi,z) = a_{1} \widetilde{LG}^{l_{1}}_{0}(\rho,\varphi,z)+a_{2} \widetilde{LG}^{l_{2}}_{0}(\rho,\varphi,z),
	\end{split}
\end{equation}
where $a_{1}$ and $a_{2}$ are complex coefficients satisfying $\vert a_{1}\vert^{2}+\vert a_{2}\vert^{2}=1 $. 

By combining Eqs. (\ref{xve}), (\ref{xcurrent}), (\ref{xLG}), and (\ref{spLG}),   the radial velocity and angular velocity in $x$-Bohm theory can be obtained as follows:
\begin{small}
	\begin{equation}\label{rave}
		\begin{split}
			& v_{r}^{x}=\frac{k\cdot r}{m} \frac{z}{z^2+z_R^2}-\dfrac{1}{m\cdot r}\dfrac{a_{1}a_{2}R_{1}R_{2}(\vert l_{1}\vert-\vert l_{2}\vert){\rm sin}(\xi_{x})}{\vert a_{1}\vert^2R_{1}^2+\vert a_{2}\vert^2R_{2}^2+2a_{1}a_{2}R_{1}R_{2}{\rm cos}(\xi_{x})},\\
			& \omega_{{\theta}}^{x} = \dfrac{-1}{m\cdot r^{{2}}}\dfrac{\vert a_{1}\vert^2R_{1}^2 l_{1}+\vert a_{2}\vert^2R_{2}^2 l_{2}+ a_{1}a_{2}R_{1}R_{2}(l_{1}+ l_{2}){\rm cos}(\xi_{x})}{\vert a_{1}^2\vert^2R_{1}^2+\vert a_{2}\vert^2R_{2}^2+2a_{1}a_{2}R_{1}R_{2}{\rm cos}(\xi_{x})},
		\end{split}
	\end{equation}
\end{small}
where $m=2\pi\hbar /(\lambda  c) $ is the mass of photon, $c $ is the speed of light, $\xi_{x}=(l_{1}-l_{2}){\theta}+ (\vert l_{1}\vert-\vert l_{2}\vert){\rm arctan}(z/z_{R})$, and $ R_{j}=\sqrt{\dfrac{2}{\pi(\vert l_{{j}}\vert)!}}\frac{1}{\omega_{z}}(\dfrac{\sqrt{2}r}{\omega_{z}})^{\vert l_{{j}} \vert}{\rm exp}(\dfrac{-r^{2}}{\omega_{z}^{2}})$ is the amplitude of $ LG^{l_{j}}_{0}(r,{\theta},z) $,  $ j\in\lbrace 1,2 \rbrace $. Consequently, the position trajectory can be expressed as
$ r \cdot \textbf{e}_r + \theta \cdot \textbf{e}_{\theta} $, where
\begin{equation}\label{xtLG}
	\begin{split}
		& r  = r_0+\int_0^z v_r^x / \sqrt{c^2-(v_r^x)^2-(\omega_\theta^x \cdot r)^2} dz, \\
		&\theta  = \theta_0+\int_0^z \omega_\theta^x /\sqrt{c^2-(v_r^x)^2-(\omega_\theta^x \cdot r)^2} dz.\\	
	\end{split}
\end{equation}
Additionally, by tracking  $ p_r^{{x}} =m \cdot v_{r}^{x} $ and ${ p_{\theta}^x  = m \cdot (\omega_\theta^x \cdot r)} $ as $z$  evolves, one can obtain the momentum trajectories $ p_r^{{x}}  \cdot \textbf{e}_r + p_{\theta}^x  \cdot \textbf{e}_{\theta} $ in  $x$-Bohm theory.

Since the LG beam propagates in free space, i.e., the potential energy function $V(r,\theta,z)=0$, the wave function in momentum space remains unchanged with respect to propagate distance. This implies that both the radial velocity {$v_\rho^p $} and angular velocity {$\omega_\varphi^p $} are equal to zero during the propagation, and the corresponding momentum remains unchanged, i.e., {$\rho  = \rho_0$ and $\varphi  = \varphi_0$}. However, it can be deduced from Eqs. (\ref{xptra}) and (\ref{spLG}) that the position trajectory in $p$-Bohm theory changes as follows 
\begin{small}
	\begin{equation}\label{ptLG}
		\begin{split}
			& {x_\rho^p }= 2\pi^{2}\omega_{0}^{2}\rho \dfrac{z}{z_{r}}+\dfrac{1}{\rho}\dfrac{a_{1}a_{2}R'_{1}R'_{2}(\vert l_{1}\vert-\vert l_{2}\vert){\rm sin}(\xi_{p})}{\vert a_{1}\vert^2 R_{1}^{'2}+\vert a_{2}\vert^2R_{2}^{'2}+2a_{1}a_{2}R'_{1}R'_{2}{\rm cos}(\xi_{p})},\\
			& {x_\varphi^p } = \dfrac{1}{\rho}\dfrac{\vert a_{1}\vert^2R_{1}^{'2} l_{1}+\vert a_{2}\vert^2R_{2}^{'2} l_{2}+ a_{1}a_{2}R'_{1}R'_{2}(  l_{1}+ l_{2} ){\rm cos}(\xi_{p})}{\vert a_{1}^2\vert^2R_{1}^{'2}+\vert a_{2}\vert^2R_{2}^{'2}+2a_{1}a_{2}R'_{1}R'_{2}{\rm cos}(\xi_{p})},
		\end{split}
	\end{equation}
\end{small}
where $ \xi_{p}=(l_{1}-l_{2})\varphi+ (\vert l_{1}\vert-\vert l_{2}\vert)\pi/2$ and $ R'_{j}=\sqrt{\dfrac{2\pi}{ \vert l_{j}\vert !}} \omega_{0} ( \sqrt{2}\pi\omega_{0}\rho)^{\vert l_{j} \vert}{\rm exp}(-\pi^{2}\rho^{2}\omega_{0}^{2})$ is the  amplitude of $ \widetilde{ LG}^{l_{j}}_{0}(\rho,\varphi,z) $, and $ j\in\lbrace 1,2 \rbrace $. Similar to $x$-Bohm theory, the radial velocity $v_{x\rho}^p$ and angular velocity $\omega_{x\varphi}^p$ in position space of  $p$-Bohm theory can be defined as 
\begin{equation}\label{vpxLG}
	\begin{split}
		v_{x\rho}^p = \dfrac{\rho}{m},\quad
		\omega_{x\varphi}^p = \dfrac{\varphi}{m \cdot x_\rho^p}.
	\end{split}
\end{equation}

\section{Visualize the propagation trajectory of LG beam}
To provide a more intuitive representation of the propagation characteristics of different LG beams, we numerically reconstruct the three-dimensional position and momentum trajectories for both single-mode LG beams and their superpositions. In the subsequent calculations, the parameters used are $\lambda = 810$ nm and $w_0=0.5$ mm. The corresponding code can be found in \cite{github} .

\subsection{The results for a single-mode LG beam}\label{subsec2}


\begin{figure}[!htbp]
	\centering\includegraphics[width=9cm]{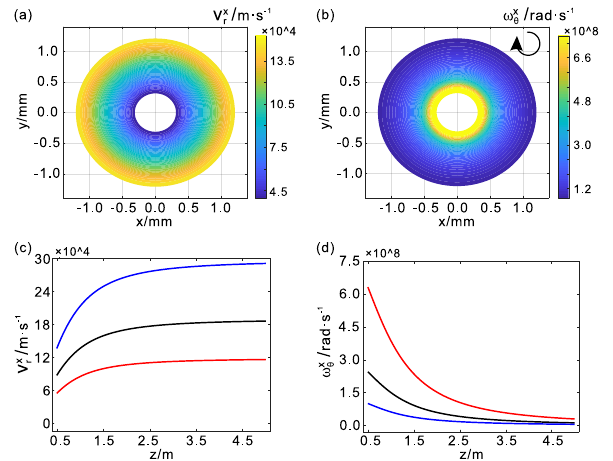}
	\caption{The velocity of a single-mode LG beam in $x$-Bohm theory with $l_{1}=3$ and $a_{1}=1$. (a) and (b) show the radial velocity and angular velocity in the $ z=0.5$ m plane, respectively. At each point, the direction of angular velocity is the same, as indicated in the upper right corner of (b). (c) and (d) show how the radial velocity and angular velocity change with the propagation distance, respectively.
	}
	\label{sLGx}
\end{figure}


\begin{figure*}[!htbp]
	\centering\includegraphics[width=13cm]{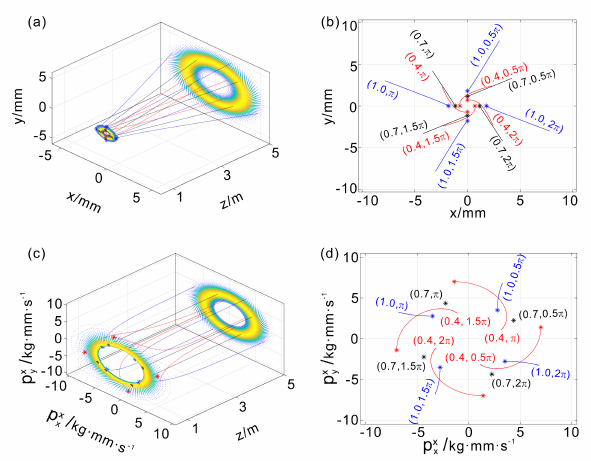}
	\caption{The trajectory of a single-mode LG beam in $x$-Bohm theory with $l_{1}=3$ and $a_{1}=1$. (a) shows the three-dimensional position trajectory of the LG beam from $ z=0.5 $ m to $ z=5 $ m. (b) shows the projection of (a) on the xoy plane. (c) shows the three-dimensional momentum trajectory of the LG beam from $ z=0.5 $ m to $ z=5$ m. (d) shows the projection of (c) on the xoy plane. $*$ remarks the starting point. Lines are labeled by initial polar coordinates.  
	}
	\label{sLGx_2}
\end{figure*}

First, we investigate the trajectory of a single-mode LG beam in $x$-Bohm theory with  $l_{1}=3$ and $a_{1}=1$. Fig. \ref{sLGx}(a) and Fig. \ref{sLGx}(b) show the radial velocity and angular velocity of the LG beam in the $z=0.5$ m plane. It can be seen that both velocities depend only on the off-axial distance (i.e., radius); with the distance decreasing, the radial velocity decreases, while the angular velocity increases. This implies that the closer to the center of the LG beam,  the slower the divergence and the faster the rotation. Additionally, as depicted in Fig. \ref{sLGx}(c) and Fig. \ref{sLGx}(d), both velocities change quickly in the near field and become constant in the far field. With the help of Eqs. (\ref{xve}) and (\ref{xxtra}), we can further derive the three-dimensional position trajectory. Four starting positions are chosen at each radius, and their corresponding three-dimensional position trajectories from $z = 0.5$ m to $z =5$ m are shown in Fig. \ref{sLGx_2}(a). Obviously, the rotation velocity of the trajectory decreases as the propagation distance increases. And, inner trajectories (red curves) exhibit faster rotation and slower divergence, while outer trajectories (blue curves) show the opposite trend, consistent with the observed changes in velocity. See Fig. \ref{sLGx_2}(b), the top view of Fig. \ref{sLGx_2}(a) for more clarity. It should be noted that the observed divergence and  rotation  will induce a reduction in the longitudinal propagation velocity. Consequently, in the communication task, information embedded in LG beams with different initial off-axial distances and propagation distances does not arrive simultaneously, necessitating corrections  \cite{BHM2016,LRW2018}.

By tracking the momentum of each position, the corresponding momentum trajectory can be obtained, as illustrated in Fig. \ref{sLGx_2}(c) and Fig. \ref{sLGx_2}(d). While the intensity distribution of the LG beam in momentum space retains a ring-shaped pattern without significant expansion, the momentum trajectories are not flat lines, excluding those starting at peak intensity (black markers). For trajectories starting within the peak intensity, the radial velocity is positive, and the angular velocity is negative, leading to inward convergence and clockwise rotation. Conversely, for trajectories starting outside the peak intensity, the radial velocity is negative, and the angular velocity is positive, resulting in outward divergence and counter-clockwise rotation. Counter-intuitively, these trajectories ``intersect" each other during propagation.


\begin{figure}[!htbp]
	\centering\includegraphics[width=9cm]{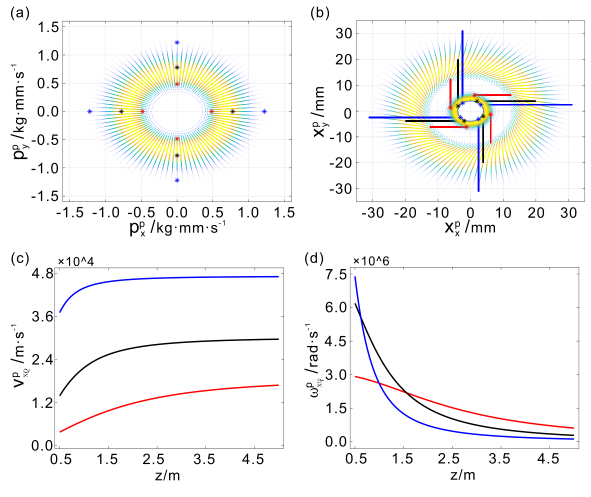}
	\caption{The results of a single-mode LG beam in $p$-Bohm theory with $l_{1}=3$ and $a_{1}=1$. (a) and (b) respectively show the projections of the three-dimensional momentum  and position trajectories in the $z=0.5$ m plane. The  background density plots in (a) and (b) represent the intensity distribution in the respective planes. (c) and (d) show how the radial velocity and angular velocity change with the propagation distance, respectively.  In (a)-(d), lines with the same color have the same radial coordinate.  
	}
	
	\label{sLGp}
\end{figure}


\begin{figure*}[!htbp]
	\centering\includegraphics[width=15cm]{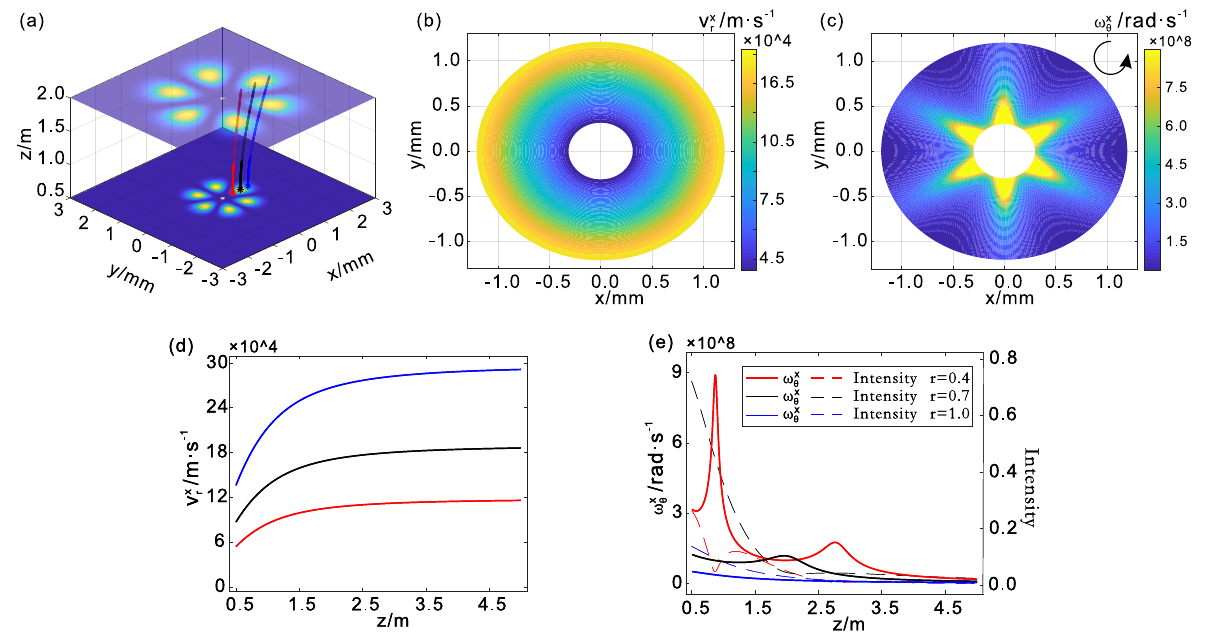}
	\caption{The results of the superposed-mode LG beam in $x$-Bohm theory with  $l_1 = -l_2 = 3$ and (a) $a_1 = a_2 = 1/\sqrt{2}$, (b-e) $a_1 = 1/\sqrt{10}$, $a_2 = 3/\sqrt{10}$. (a) shows the three-dimensional position trajectory from $ z=0.5 $ m to $ z=2 $ m. (b) and (c) show the radial velocity and angular velocity distributions in the $ z=0.5 $ m plane, respectively. At each point, the direction of angular velocity is the same, as indicated in the upper right corner of (c). (d) and (e) show how the radial velocity and angular velocity change with the propagation distance, respectively. In (a) and (d)-(e), lines with the same color have the same radial coordinate. The background density plots in (a) represents the intensity distribution in the respective plane. }
	\label{tLGx}
\end{figure*}
On the other hand, for the freely propagating LG beam, its intensity distribution in momentum space remains constant. Consequently, both the radial velocity {$v_\rho^p $} and angular velocity {$\omega_\varphi^p $} are equal to zero during the propagation. The momentum trajectories in $p$-Bohm theory are represented by straight lines, specifically given by $\rho = \rho_0$ and $\varphi = \varphi_0$  (see Fig. \ref{sLGp}(a)). In contrast to the results depicted in Fig. \ref{sLGx_2}(c) and Fig. \ref{sLGx_2}(d), these trajectories exhibit no divergence and rotation. By tracking each position along the momentum trajectory, the corresponding position trajectory can be obtained, as shown in Fig. \ref{sLGp}(b). Surprisingly, these $p$-Bohm position trajectories originate from positions with vanishingly low probabilities according to $x$-Bohm. Notably, in $p$-Bohm theory, position trajectories with the same initial angular coordinate divergence in parallel. This divergence arises from the fact that, while the radial velocity in the $p$-Bohm theory undergoes a similar evolution to that in the $x$-Bohm theory, the angular velocity, particularly in the near field, follows a different trend, being inversely proportional to the radius. See Fig. \ref{sLGp}(c) and (d) for more clarity.

\subsection{The results for superposed-mode LG beams}\label{subsec2}

In this section, we consider the case when $a_2\neq 0$, which indicates the beam is in a superposition state of two LG beams with different $l_1 $ and $l_2$.  The intensity distribution of the superposed-mode LG beam is petal-shaped, with the number of petals being $\vert l_2-l_1 \vert$. It is straightforward to deduce from Eq. (\ref{rave}) that when $l_1 = -l_2$ and $a_1 = a_2$, the angular velocity becomes zero. As shown in Fig. \ref{tLGx}(a), the superposed-mode LG beam exhibits rotation-free propagation, yielding seemingly trivial characteristics.


\begin{figure}[!htbp]
	\centering\includegraphics[width=9cm]{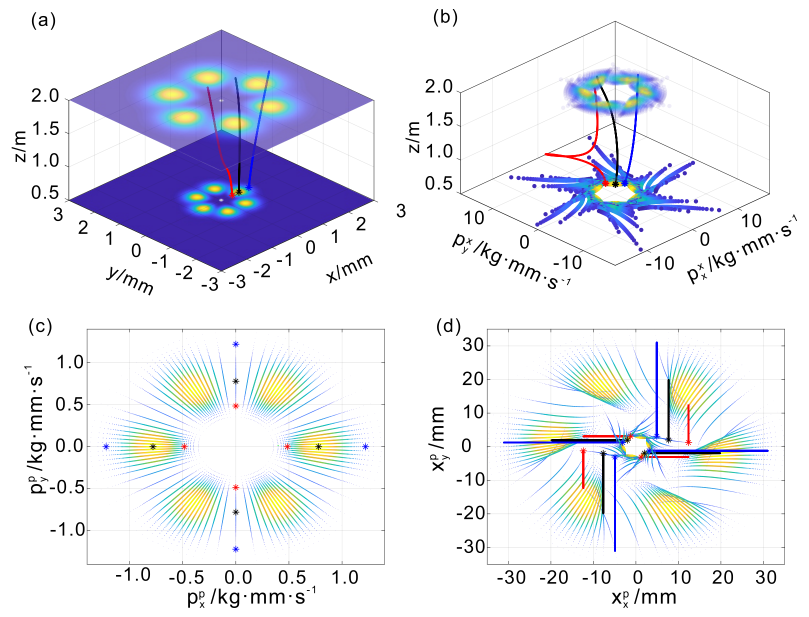}
	\caption{The trajectory of the superposed-mode LG beam with  $l_1 = -l_2 = 3$ and $a_1 = 1/\sqrt{10}$, $a_2 = 3/\sqrt{10}$.  (a) and (b) show the three-dimensional position and momentum trajectory of the LG beam from $ z=0.5 $ m to $ z=2 $ m in $x$-Bohm theory, respectively. The momentum trajectory and position trajectory of the superposed-mode LG beam in $p$-Bohm theory  are shown in (c) and (d), respectively. 
	In (a-d), lines with the same color have the same radial coordinate. The background density plots in (a-d)  represents the intensity distribution in the respective plane.  }
	\label{tLGx_2}
\end{figure}

To better illustrate the distinction between superposed-mode and single-model LG beams, we set $l_1=-l_2 = 3$, $a_1=1/\sqrt{10}$  and $a_2=3/\sqrt{10}$ to investigate the propagation characteristics of the superposed-mode LG beam. The corresponding results in $x$-Bohm theory are presented in  Fig. \ref{tLGx}(b-e) and Fig. \ref{tLGx_2}(a-b) . As shown in Fig. \ref{tLGx}(b) and Fig. \ref{tLGx}(d), the change in the radial velocity is similar to that of a single-mode LG beam. However, due to interference, the angular velocity exhibits periodic oscillation. Fig. \ref{tLGx}(c) and Fig. \ref{tLGx}(e) show that in regions with lower angular velocity,  trajectories converge together, forming bright petals; conversely, in regions with higher angular velocity, trajectories disperse and form dark petals. After integrating the velocity, the $x$-Bohm position trajectory can be obtained. Fig. \ref{tLGx_2}(a)  clearly shows that the inner trajectory moves from one bright petal to another. Furthermore, by tracking the momentum at each point on this trajectory, the momentum trajectory in $x$-Bohm theory can be determined. As shown in Fig. \ref{tLGx_2}(b), the radial velocity in non-primary space is not always positive, especially for the inner trajectory indicated by the red line. Comparing Fig. \ref{tLGx_2}(a) and Fig. \ref{tLGx_2}(b), a distinctive observation emerges: while primary space consistently maintains a perfect petal-shaped intensity distribution, non-primary space can only achieve it in the far field. In $p$-Bohm theory, the petal-shaped intensity distribution remains unchanged in the primary space. However, in the non-primary space, it diverges with the increase of the propagation distance. Interestingly, the evolutions of the $p$-Bohm momentum and position trajectories are similar to those of the single-mode LG beam, yet they deviate from the trajectory observed in $x$-Bohm theory. See Fig. \ref{sLGp}(a-b), Fig. \ref{tLGx_2}(a-b), and Fig. \ref{tLGx_2}(c-d)  for more clarity.

\begin{figure}[!htbp]
	\centering\includegraphics[width=9cm]{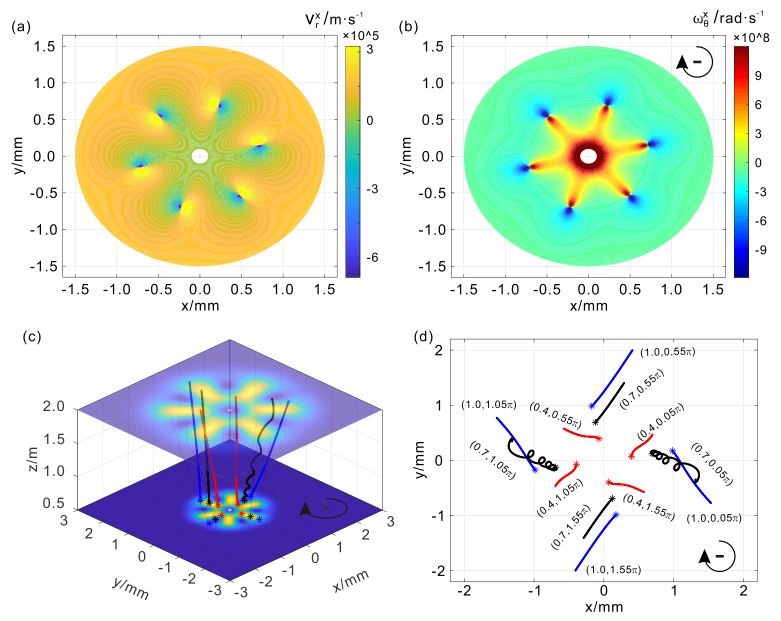}
	\caption{The results of superposed-mode LG beam in the $x$-Bohm theory with $l_1 =-1$, $l_2 =5$ and  $a_1 = a_2 = 1/\sqrt{2}$. (a) and (b) show the radial velocity and angular velocity in the $z = 0.5$ m plane. (c) shows the three-dimensional position trajectories from $z = 0.5$ m to $z =2$ m. Red and blue trajectories respectively start at the extreme intensity of the single-mode LG beam with $ l_{1}=-1 $ and $ l_{1}=5 $, while black trajectories start at that of their superposition. The background density plots represent the intensity distribution in the respective planes.  (d) shows the projection of (c) on the xoy plane. $*$ remarks the starting point.}
	\label{tLGp}
\end{figure}

The analysis above highlights that the $x$-Bohm theory offers a more familiar dynamical scenario. To comprehensively investigate the propagation characteristics of superposed-mode LG beams, we further explore the dynamic behaviour of LG beam with $l_1 =-1$, $l_2 =5$ and  $a_1 = a_2 = 1/\sqrt{2}$ within the framework of $x$-Bohm theory. Fig. \ref{tLGp}(a) and Fig. \ref{tLGp}(b) show that both the radial velocity and angular velocity change direction around the center of the  the dark intensity patterns. The three-dimensional position trajectories from $z = 0.5$ m to $z =2$ m are shown in Fig. \ref{tLGp}(c). The red and blue trajectories respectively start at the extreme intensity of the single-mode LG beam with $ l_{1}=-1 $ and $ l_{1}=5 $, while the black trajectories start at that of their superposition.  Obviously, the red trajectory rotates in the opposite direction as the superposed-mode LG beam, while both the blue and black trajectories rotate in the same direction. Additionally, the black trajectory, starting from the minimum intensity, rotates the most quickly. See Fig. \ref{tLGp}(d), the top view of Fig. \ref{tLGp}(c) for more clarity.

\section{Conclusions and Discussion}

In this work, we construct the three-dimensional trajectory of LG beams employing the $x$-Bohm theory and $p$-Bohm theory. Our results reveal that the divergence of a single-mode LG beam increases with both off-axis distance and propagation distance, while the rotation behavior displays a distinct dynamic. We rigorously quantify the divergence velocity and rotation velocity, allowing precise adjustment of the arrival time of encoded information in LG beams, which is crucial for communication tasks. Additionally, we find that the distinctive petal-shaped intensity distribution in the superposed-mode LG beam emerges as the trajectories converge when approaching the maximum intensity and diverge on coming out of it. That is why the energy concentrates in the ``rear"  and dissipates near the ``front" of the bright petal. Of course, all these processes can be described in the optical wave language which is used in the original works, but the employment of the propagation trajectory makes the behavior physically clear. In addition, when the absolute values of the angular momentum indices in the superposed-mode LG beam are different, the divergence and rotation direction of the trajectory do not always align with that of the intensity distribution. Instead, they depend on the initial position and propagation distance.  Intriguingly, discrepancies are observed between the position and momentum trajectories in $x$-Bohm theory and their counterparts in $p$-Bohm theory. This reveals an asymmetry between position and momentum and underscores the importance of selecting a primary ontological variable \cite{PFG2020}.

\begin{backmatter}
\bmsection{Funding} This work was supported by the Fundamental Research Funds for the Central Universities (Grant No. 202364008), the Natural Science Foundation of Shandong Province of China (Grant No. ZR2021ZD19), and the Young Talents Project at Ocean University of China (Grant No. 861901013107).

\bmsection{Data availability} The data sets generated during and/or analyzed during the current study are available from the corresponding author on reasonable request.

\bmsection{Disclosures} The authors have no competing interests to declare that are relevant to the content of this article.

\bmsection{Disclosures} The authors promise that the results of this article are correct and original and it has not been published in other journals before.

\end{backmatter}


\ifthenelse{\equal{\journalref}{aop}}{%
\section*{Author Biographies}
\begingroup
\setlength\intextsep{0pt}
\begin{minipage}[t][6.3cm][t]{1.0\textwidth} 
  \begin{wrapfigure}{L}{0.25\textwidth}
    \includegraphics[width=0.25\textwidth]{john_smith.eps}
  \end{wrapfigure}
  \noindent
  {\bfseries John Smith} received his BSc (Mathematics) in 2000 from The University of Maryland. His research interests include lasers and optics.
\end{minipage}
\begin{minipage}{1.0\textwidth}
  \begin{wrapfigure}{L}{0.25\textwidth}
    \includegraphics[width=0.25\textwidth]{alice_smith.eps}
  \end{wrapfigure}
  \noindent
  {\bfseries Alice Smith} also received her BSc (Mathematics) in 2000 from The University of Maryland. Her research interests also include lasers and optics.
\end{minipage}
\endgroup
}{}

\end{document}